\documentclass[pre,twocolumn,showpacs]{revtex4}
\usepackage{graphicx}

\newcommand{\beq}{\begin{equation}}
\newcommand{\eeq}{\end{equation}}
\newcommand{\beqa}{\begin{eqnarray}}
\newcommand{\eeqa}{\end{eqnarray}}
\newcommand{\ba}{\begin{array}}
\newcommand{\ea}{\end{array}}

\begin{document}

\title{Vector solitons in nearly-one-dimensional Bose-Einstein condensates}
\author{Luca Salasnich$^{1}$ and Boris A. Malomed$^{2}$}
\affiliation{$^1$CNISM and CNR-INFM, Unit\`a di Padova, 
Dipartimento di Fisica ``Galileo
Galilei'', Universit\`a di Padova, Via Marzolo 8, 35131 Padova, Italy \\
$^{2}$Department of Interdisciplinary Studies, School of Electrical
Engineering, Faculty of Engineering, Tel Aviv University, Tel Aviv 69978,
Israel}

\begin{abstract}
We derive a system of nonpolynomial Schr\"{o}dinger equations (NPSEs) for 
one-dimensional wave functions of two components in a binary self-attractive 
Bose-Einstein condensate loaded in a cigar-shaped trap. The system is 
obtained by means of the variational approximation, starting from the 
coupled 3D Gross-Pitaevskii equations and assuming, as usual, the 
factorization of 3D wave functions. The system can be obtained in a
tractable form under a natural condition of symmetry between the two
species. A family of vector (two-component) soliton solutions is
constructed. Collisions between orthogonal solitons (ones belonging to the
different components) are investigated by means of simulations. The
collisions are essentially inelastic. They result in strong excitation of
intrinsic vibrations in the solitons, and create a small orthogonal
component (``shadow") in each colliding soliton. The collision may initiate
\textit{collapse}, which depends on the mass and velocities of the solitons.
\end{abstract}

\maketitle


\section{Introduction}

Bose-Einstein condensates (BECs) with attractive interactions between atoms
can form stable wave packets in nearly one-dimensional (1D) ``cigar-shaped"
traps, which provide for tight confinement in two transverse directions,
while leaving the condensate almost free along the longitudinal axis. This
trapping geometry made it possible to create stable bright solitons \cite%
{Khaykovich02} and trains of such solitons \cite{Strecker02} in the $^{7}$Li
condensate, in which the interaction between atoms was made weakly
attractive by means of the Feshbach-resonance technique. In the $^{85}$Rb
condensate trapped under similar conditions, stronger attraction between
atoms leads to controllable collapse and creation of nearly 3D solitons \cite%
{Cornish}.

This experimentally relevant situation is described by effective 1D
equations which may be derived from the full 3D Gross-Pitaevskii equation
(GPE) under various conditions and by means of different approximations \cite%
{PerezGarcia98}-\cite{Napoli}. In some cases, the deviation of the effective
equation from a straightforward 1D variant of the GPE amounts to keeping an
extra self-attractive \emph{quintic} term in the equation, which may be
sufficient to essentially alter properties of the corresponding solitons
\cite{Shlyap02,Brand06,Lev}. A more consistent derivation, that starts with
the factorization of the 3D wave function into the product of a transverse
one (it actually represents the ground state of the 2D harmonic oscillator)
and arbitrary slowly varying longitudinal (one-dimensional) wave function,
and then uses the variational approximation \cite{Progress}, leads to a more
sophisticated but also more accurate \textit{nonpolynomial Schr\"{o}dinger
equation} (NPSE) for the longitudinal wave function \cite{sala1,sala2}. The
above-mentioned simplified equation featuring the combination of cubic and
quintic terms can be then obtained by an expansion of the NPSE for the case
of a relatively weak nonlinearity \cite{Lev}. 
The ratio of the coefficients in front of the cubic and quintic terms
in the model derived in Ref. \cite{Shlyap02} is not the 
same as follows from the expansion of the NPSE, which is explained 
by a coarser character of the approximation used 
in that work (the approximation did not allow a deviation 
of the nonlinearity different from the cubic-quintic form). 

A physically significant generalization of the above-mentioned equations is
a system of two nonlinearly coupled equations for a binary BEC, which can be
created in the experiment by means of the \textit{sympathetic-cooling}
technique \cite{binary}. Accordingly, a relevant problem is to derive a
system of effective 1D equations for a mixture of two BEC species in the
cigar-shaped trap, starting from the two coupled GPEs in the 3D space. In
this work, we aim to derive such a system in the form of coupled NPSEs,
using a generalized version of the method elaborated in Refs. \cite%
{sala1,sala2,sala3}.

The paper is organized as follows. The derivation of the coupled NPSE
system, which is based on the variational approximation, is presented in
Section II. In the most general case, it leads to a cumbersome system.
However, we demonstrate that, under a natural condition of a symmetry
between the two species, the equations may be reduced a tractable closed
system of two NPSEs for longitudinal wave functions of the two components.
Then, in Section III, we consider solutions for vector solitons (i.e.,
two-component ones) generated by this system; the solutions are found in an
implicit analytical form up to a point where they cease to exist due to
collapse.

A natural application of the thus derived NPSE system is to consider
collisions between two \textit{orthogonal} solitons, which belong to the two
different components. This analysis, based on numerical simulations, is
reported in Section IV. The collisions are inelastic, which is manifested in
the excitation of intrinsic oscillations in the solitons after the
collision, and generation of small ``shadows" in them (each soliton captures
and keeps a small share of atoms from the other species). The strongest
manifestation of the inelasticity, as we demonstrate in Section IV, is a
possibility to initiate collapse by the collision between two orthogonal
solitons (which depends on their relative velocity). The paper is concluded
by Section V.

\section{Coupled nonpolynomial Schr\"{o}dinger equations}

The system of 3D GPEs for a dilute binary condensate, confined in the
transverse direction by a strong harmonic potential with frequency $\omega
_{\bot }$ and in the axial direction by a generic weak potential $V(z)$, can
be derived from the Lagrangian density,
\[
L=\sum_{k=1,2}\psi _{k}^{\ast }\left[ i\,\partial _{t}+{\frac{1}{2}}\nabla
^{2}-{\frac{1}{2}}(x^{2}+y^{2})\right.
\]%
\begin{equation}
\left. -V(z)-{\pi \,g_{k}}|\psi _{k}|^{2}\right] \psi _{k}-2\pi
\,g_{12}|\psi _{1}|^{2}|\psi _{2}|^{2}\;.  \label{L0}
\end{equation}%
Here $\psi _{k}(\mathbf{r},t)$ is the macroscopic wave function of the $k$%
-th species, which is subjected to the normalization condition,
\begin{equation}
\int \int \int \left\vert \psi _{k}(x,y,z)\right\vert ^{2}dxdydz=N_{k}\;,
\label{norm}
\end{equation}%
where $N_{k}$ is the number of atoms in the $k$-th species, and
\begin{equation}
g_{k}\equiv 2a_{k}/a_{\bot },~g_{12}\equiv 2a_{12}/a_{\bot }  \label{ggg}
\end{equation}%
are strengths of the intra- and inter-species interactions, where $a_{k}$
and $a_{12}$ are the scattering lengths, and $a_{\bot }=\sqrt{\hbar
/(m\omega _{\bot })}$ the transverse harmonic-confinement length. Here we
consider the binary condensate with attraction between atoms, which implies
that both $a_{1,2}$ and $a_{12}$ are \emph{negative}. In the Lagrangian
density, lengths, time, and energy are written in units $a_{\bot }$, $\omega
_{\bot }^{-1}$, and $\hbar \omega _{\bot }$, respectively.

The ordinary variational procedure applied to Eq. (\ref{L0}) gives rise to
the coupled 3D\ GPEs,
\[
i\,\partial _{t}\psi _{k}=\left[ -{\frac{1}{2}}\nabla ^{2}+{\frac{1}{2}}%
(x^{2}+y^{2})+V(z)\right.
\]%
\begin{equation}
\left. +2\pi \,g_{k}|\psi _{k}|^{2}+2\pi \,g_{12}|\psi _{3-k}|^{2}\right]
\psi _{k},~k=1,2.  \label{GPE}
\end{equation}
Our objective here is to derive a system of effective 1D NPSEs, following
the lines of the derivation of the NPSE for the single-component condensate
developed in Ref. \cite{sala1} [its generalization for an axially nonuniform
trapping potential, with $\omega _{\perp }=\omega _{\perp }(z)$, was
reported in Ref. \cite{Napoli}]. Using the cylindrical coordinates $\left(
r,\theta \right) $ in the transverse plane $\left( x,y\right) $, we adopt
the usual ansatz for the wave functions strongly localized in this plane,
and weakly confined in the axial direction, $z$:%
\begin{equation}
\psi _{k}(r,z,t)={\frac{1}{\sqrt{\pi }\sigma _{k}(z,t)}}\exp {\left\{ -{%
\frac{r^{2}}{2\sigma _{k}(z,t)^{2}}}\right\} }\,f_{k}(z,t)\;,  
\label{ansatz}
\end{equation}%
where real $\sigma _{k}(z,t)$ and complex $f_{k}(z,t)$ are dynamical fields,
the latter ones obeying normalization $\int_{-\infty }^{+\infty }\left\vert
f_{k}(z)\right\vert ^{2}dz=N_{k}$, as it follows from Eqs. (\ref{norm}) 
and (\ref{ansatz}).

Inserting this ansatz in Lagrangian density (\ref{L0}), performing the
integration in the transverse plane, and neglecting derivatives of 
$\sigma(z,t)$ (for the same reasons as in Refs. \cite{sala1,Napoli}), one can
derive the following effective Lagrangian:
\[
{\bar{L}}=\sum_{k=1,2}f_{k}^{\ast }\left[ i\,\partial _{t}+{\frac{1}{2}}%
\partial _{z}^{2}-{\frac{1}{2}}\left( \frac{1}{{\sigma }_{k}^{2}}+\sigma
_{k}^{2}\right) -V(z)\right.
\]%
\[
\left. -{\frac{1}{2}}{\frac{g_{k}}{\sigma _{k}^{2}}}|f_{k}|^{2}\right]
f_{k}-2{\frac{g_{12}|f_{1}|^{2}|f_{2}|^{2}}{(\sigma _{1}^{2}+\sigma _{2}^{2})%
}}\;.
\]%
This Lagrangian gives rise to a system of four Euler-Lagrange equations,
obtained by varying ${\bar{L}}$ with respect to $f_{k}^{\ast }$ and $\sigma
_{k}$:
\[
i\,\partial _{t}f_{k}=\left[ -{\frac{1}{2}}\partial _{z}^{2}+V(z)+{\frac{1}{2%
}}\left( {\frac{1}{\sigma _{k}^{2}}}+\sigma _{k}^{2}\right) \right.
\]%
\begin{equation}
\left. +{\frac{g_{k}}{\sigma _{k}^{2}}}|f_{k}|^{2}+2{\frac{g_{12}}{(\sigma
_{1}^{2}+\sigma _{2}^{2})}}|f_{3-k}|^{2}\right] f_{k}\;,  \label{sala-npse}
\end{equation}%
\begin{equation}
\sigma _{k}^{4}=1+{g_{k}|f_{k}|^{2}}+{4g_{12}|f_{3-k}|^{2}}{\frac{\sigma
_{k}^{4}}{(\sigma _{1}^{2}+\sigma _{2}^{2})^{2}}}\;,  \label{sala-sigma}
\end{equation}%
with $k=1,2$ in both equations (\ref{sala-npse}) and (\ref{sala-sigma}).
This is a full system of the coupled NPSEs describing the two-component
nearly-1D BEC. 

Note that the ansatz (\ref{ansatz}) is relevant when the transverse 
confinement size is much smaller than a characteristic axial length 
of a structure (in particular, soliton) to be obtained as a solution 
of the axial equation. Physically, this means that the quantum pressure 
in the transverse direction is much stronger 
than the nonlinear self-attraction in the condensate. 

\section{Vector bright solitons}

In the subsequent analysis of the coupled NPSEs, we focus on the most
natural symmetric case, when the (negative) effective nonlinearity
coefficients accounting for the intra- and inter-species interactions are
equal, namely,
\begin{equation}
g_{12}=g_{1}=g_{2}\equiv g.  \label{g}
\end{equation}%
As follows from Eqs. (\ref{ggg}), this relation takes place, in particular,
when the all scattering lengths are equal. In the symmetric case, we assume
that numbers of atoms in the two species are equal too. We note that, for
the self-repulsive binary BEC, with $g>0$ (recall here we are going to
consider the case of $g<0$), Eq. (\ref{g}) may pose a formal problem, as it
precisely corresponds to the miscibility-immiscibility threshold in the
infinite system. Nevertheless, the problem does not really take place, as
the pressure exerted by the external potential shifts the equilibrium
towards the miscibility (see, e.g., Ref. \cite{boris0}), hence the case
corresponding to relation (\ref{g}) in the repulsive binary BEC is not going
to be a degenerate (i.e., structurally unstable) one.

If constraint (\ref{g}) holds, Eqs. (\ref{sala-sigma}) take the form
\begin{equation}
\sigma _{k}^{4}=1+g|f_{k}|^{2}+4g|f_{3-k}|^{2}\frac{\sigma _{k}^{4}}{\left(
\sigma _{1}^{2}+\sigma _{2}^{2}\right) ^{2}}\;,  \label{sigma}
\end{equation}%
and admit an \emph{exact symmetric solution}:
\begin{equation}
\sigma _{1}^{2}=\sigma _{2}^{2}=\sqrt{1+g\left(
|f_{1}|^{2}+|f_{2}|^{2}\right) }\equiv \sigma _{0}^{2}\;.  \label{equal}
\end{equation}%
Of course, there remains a question if some additional \emph{asymmetric}
solutions to Eqs. (\ref{sigma}) may also exist. One may assume that an
asymmetric solution, if any, branches off from the symmetric one through a
\textit{bifurcation}. Then, close to the bifurcation point, one will have $%
\sigma _{1,2}^{2}=\sigma _{0}^{2}\left( 1+\delta _{1,2}\right) $, with some
infinitesimal $\delta _{1}\neq \delta _{2}$ [$\sigma _{0}$ is the symmetric
solution given by Eq. (\ref{equal})]. Substituting this in Eqs. (\ref{sigma}%
) and linearizing them in $\delta _{1}$ and $\delta _{2}$, one arrives at a
system
\begin{eqnarray}
2\delta _{k} &=&F_{k}\left( \delta _{k}-\delta _{3-k}\right) ,\;\;k=1,2\;,
\label{linear} \\
F_{k} &\equiv &g|f_{k}|^{2}/\sigma _{0}^{4}\;.  \label{F}
\end{eqnarray}%
The resolvability condition for linear system (\ref{linear}) (equating its
determinant to zero) takes the following form, after simple calculations: $%
F_{1}+F_{2}=2$. However, this condition \emph{cannot hold} for the
attractive BEC, with $g<0$, because expressions $F_{1}$ and $F_{2}$, as
given by Eq. (\ref{F}), are negative in this case. This means the
bifurcation giving rise to asymmetric solutions is impossible in the
attractive binary condensate [provided that constraint (\ref{g}) is valid],
which substantiates the use of symmetric solution (\ref{equal}).

The substitution of Eq. (\ref{equal}) in Eqs. (\ref{sala-npse}) leads to
closed-form equations for the complex amplitude functions, $f_{1}$ and $f_{2}
$,
\[
i\frac{\partial f_{k}}{\partial t}=\left[ -\frac{1}{2}\frac{\partial ^{2}}{%
\partial z^{2}}+V(z)+{g}\frac{|f_{1}|^{2}+|f_{2}|^{2}}{\sqrt{1+g\left(
|f_{1}|^{2}+|f_{2}|^{2}\right) }}\right.
\]%
\begin{equation}
\left. +\frac{1}{2}\left( {\frac{1}{\sqrt{1+g\left(
|f_{1}|^{2}+|f_{2}|^{2}\right) }}}+\sqrt{1+g\left(
|f_{1}|^{2}+|f_{2}|^{2}\right) }\right) \right] f_{k}\;.  \label{final}
\end{equation}%
Equations (\ref{final}) reduce to the familiar integrable \textit{Manakov's
system} (MS)\ \cite{Manakov},
\begin{equation}
i\,\partial _{t}f_{k}=\left[ -{\frac{1}{2}}\partial _{z}^{2}+V(z)+g\left(
|f_{1}|^{2}+|f_{2}|^{2}\right) \right] f_{k}\;,  \label{nlse}
\end{equation}%
if $g(|f_{1}|^{2}+|f_{2}|^{2})\ll 1$. Only under this condition
the system may be considered as truly one-dimensional, and only in
this limit it is integrable. Nevertheless, in the general case
Eqs. (\ref{final}) share the ``isotopic invariance" with the
Manakov's system: the nonlinearity appears solely through the
invariant combination, $\left\vert f_{1}\right\vert
^{2}+\left\vert f_{2}\right\vert ^{2}$. \ Due to this fact, Eqs.
(\ref{final}) conserve an additional dynamical invariant
(``isotopic spin"), $S=\int_{-\infty }^{+\infty }\left[
f_{1}(z)f_{2}^{\ast }(z)+f_{1}^{\ast }(z)f_{2}(z)\right] dz$, with
asterisk standing for the complex conjugate.

In the case of attraction, $g<0$, \textit{vector }(two-component)\textit{\ }%
bright solitons are looked for as $f_{k}=\exp \left( -i\mu _{k}t\right) \Phi
_{k}(z)$, where $\Phi _{1}$ and $\Phi _{2}$ are real localized functions
obeying the following coupled equations:
\[
\mu _{k}\Phi _{k}=\left[ -\frac{1}{2}\Phi _{k}^{\prime \prime }+V(z)\Phi
_{k}+{g}\frac{\Phi _{1}^{2}+\Phi _{2}^{2}}{\sqrt{1+g\left( \Phi
_{1}^{2}+\Phi _{2}^{2}\right) }}\right.
\]%
\begin{equation}
\left. +\frac{1}{2}\left( {\frac{1}{\sqrt{1+g\left( \Phi _{1}^{2}+\Phi
_{2}^{2}\right) }}}+\sqrt{1+g\left( \Phi _{1}^{2}+\Phi _{2}^{2}\right) }%
\right) \right] \Phi _{k}~.  \label{Phi}
\end{equation}%
Due to the isotopic invariance, the solitons with equal chemical potentials
of their components are tantamount to the single-component (scalar) one,
with $\Phi _{2}=0$ and $\Phi _{1}\equiv \Phi (z)$ being a solution of a
single equation,
\begin{equation}
\left[ -\frac{1}{2}\frac{d^{2}}{dz^{2}}+V(z)+\frac{1+\frac{3}{2}g\Phi ^{2}}{%
\sqrt{1+g\Phi ^{2}}}\right] \Phi =\mu \,\Phi ~.  \label{Phi0}
\end{equation}%
If a soliton solution to Eq. (\ref{Phi0}) is found, the corresponding vector
soliton can be constructed in an obvious way,
\begin{equation}
\left\{
\begin{array}{c}
f_{1}(z,t) \\
f_{2}(z,t)%
\end{array}%
\right\} =\left\{
\begin{array}{c}
\cos \theta  \\
\sin \theta
\end{array}%
\right\} \cdot \left\{
\begin{array}{c}
\exp \left( -i\mu t\right)  \\
\exp \left( -i\mu t\right)
\end{array}%
\right\} \cdot \Phi (z),  \label{theta}
\end{equation}%
with an arbitrary ``isotopic angle", $0\leq \theta \leq \pi /2$.
More general vector solitons, with different chemical potentials
in their components, are possible two. However, in the symmetric
case that we are dealing with here, it is obvious that the vector
solitons with unequal chemical potentials cannot realize the
ground state, therefore they are considered here.

For $V(z)=0$ and $g<0$, a family of soliton solutions to Eq. (\ref{Phi0})
was constructed in Refs. \cite{sala1,sala2}. In this case, setting $\Phi (z)=%
\sqrt{N}\phi (z)$, with $N_{1}=N_{2}\equiv N$, and
\begin{equation}
\gamma \equiv N|g|\;,
\end{equation}%
the field $\phi (z)$ and the chemical potential $\mu $ are given by implicit
formulas,
\[
z=\sqrt{\frac{1}{2(1-\mu )}}\mathrm{Arctanh}\left( \sqrt{\frac{\sqrt{%
1-\gamma \phi ^{2}}-\mu }{1-\mu }}\right)
\]%
\begin{equation}
-{\frac{1}{\sqrt{2}}}\sqrt{\frac{1}{1+\mu }}\tan ^{-1}\left( \sqrt{\frac{%
\sqrt{1-\gamma \phi ^{2}}-\mu }{1+\mu }}\right) \;,  \label{alb}
\end{equation}%
\begin{equation}
\gamma ={\frac{2\sqrt{2}}{3}}(2\mu +1)\sqrt{1-\mu }\;.  \label{gamma}
\end{equation}%
and the wave function satisfies the normalization condition $\int_{-\infty
}^{+\infty }\left( \phi (z)\right) ^{2}dz=1$. The family is then
characterized by the dependence of $\gamma $ on $\mu $. The inverse of Eq. (%
\ref{gamma}) demonstrates that, in terms of the $\mu (\gamma )$ dependence,
there are two branches of the soliton family, but only one of them, that
satisfies condition $d\mu /d\gamma <0$ (which is nothing else but the known
\textit{Vakhitov-Kolokolov} stability criterion \cite{VK} in the present
notation), is stable. In addition, there is a critical nonlinearity
strength, $\gamma _{c}=4/3$ (which corresponds to $\mu =1/2$), above which
the solution does not exist, because of the \textit{longitudinal collapse}
\cite{sala1,Brand04}, which is a manifestation of the 3D collapse possible
in the underlying system of GPEs, Eqs. (\ref{GPE}). In the limit of weak
nonlinearity, $\gamma \rightarrow 0$, Eq. (\ref{alb}) reduces to the
ordinary soliton waveform, $\phi (z)=\left( \sqrt{\gamma }/2\right) \mathrm{%
sech}\left( \gamma z/2\right) $ \cite{sala1,sala3}.

\section{Collisions between solitons}

A straightforward application of the system of NPSEs (\ref{final}) is to
study collisions between two \textit{orthogonal} solitons, taken in the form
of Eq. (\ref{theta}), with equal values of $\mu $ and isotopic angles $%
\theta =0$ and $\theta =\pi /2$. Using the Galilean invariance of the
equations, the velocities of the two solitons are taken to be $\pm v$. The
corresponding initial condition, at $t=0$, is
\begin{equation}
\left\{
\begin{array}{c}
\phi _{1}^{(0)}(z) \\
\phi _{2}^{(0)}(z)%
\end{array}%
\right\} =\left\{
\begin{array}{c}
e^{ivz}\phi (z-z_{0}/2) \\
e^{-ivz}\phi (z+z_{0}/2)%
\end{array}%
\right\} ,  \label{initial}
\end{equation}%
with large initial separation $z_{0}$. To determine the time-evolution of
the fields $\phi _{k}(z,t)$, $k=1,2$, we solved both NLSEs and NPSEs
numerically, by using a two-component extension of a well-tested
finite-difference code based on the Crank-Nicolson predictor-corrector
algorithm \cite{sala-numerics}. In the MS [alias the NLSE system, Eqs. (\ref%
{nlse})], which is integrable, collisions are always elastic. However, since
NPSEs (\ref{final}) are not integrable, collisions described by these
equations are expected to be \textit{inelastic}. This expectation is borne
out by Fig. 1, where we compare the collision outcomes in the MS and NPSEs
for identical sets of parameters. The figure shows the peak densities, $n_{P}
$, of both colliding solitons (which are equal, due to the symmetry of the
configuration) , as a function of time. The outcome does not depend on the
initial separation $z_{0}$ between the solitons in Eq. (\ref{initial}),
provided that it is large enough (we took $z_{0}=200$). After the collision
the MS solitons remain undisturbed (dashed lines), while their NPSE
counterparts come out from the collision with excited intrinsic oscillations
(solid lines). This result not only shows that the collision in the NPSEs is
inelastic, but also suggests that the solitons supported by this system,
i.e., ones given by Eqs. (\ref{theta}), (\ref{alb}), and (\ref{gamma}),
unlike their counterparts in the integrable MS, feature an intrinsic mode,
with a well-defined eigenfrequency, $\omega $. In fact, this mode was
predicted in Ref. \cite{sala2}, by means of the variational approximation.
It was shown that $\omega $ vanishes at $\gamma \rightarrow 0$, and it
attains a maximum close to the above-mentioned collapse threshold, $\gamma
=\gamma _{c}\equiv 4/3$.

\begin{figure}[tbp]
\centerline
{\includegraphics[height=2.35in,clip] {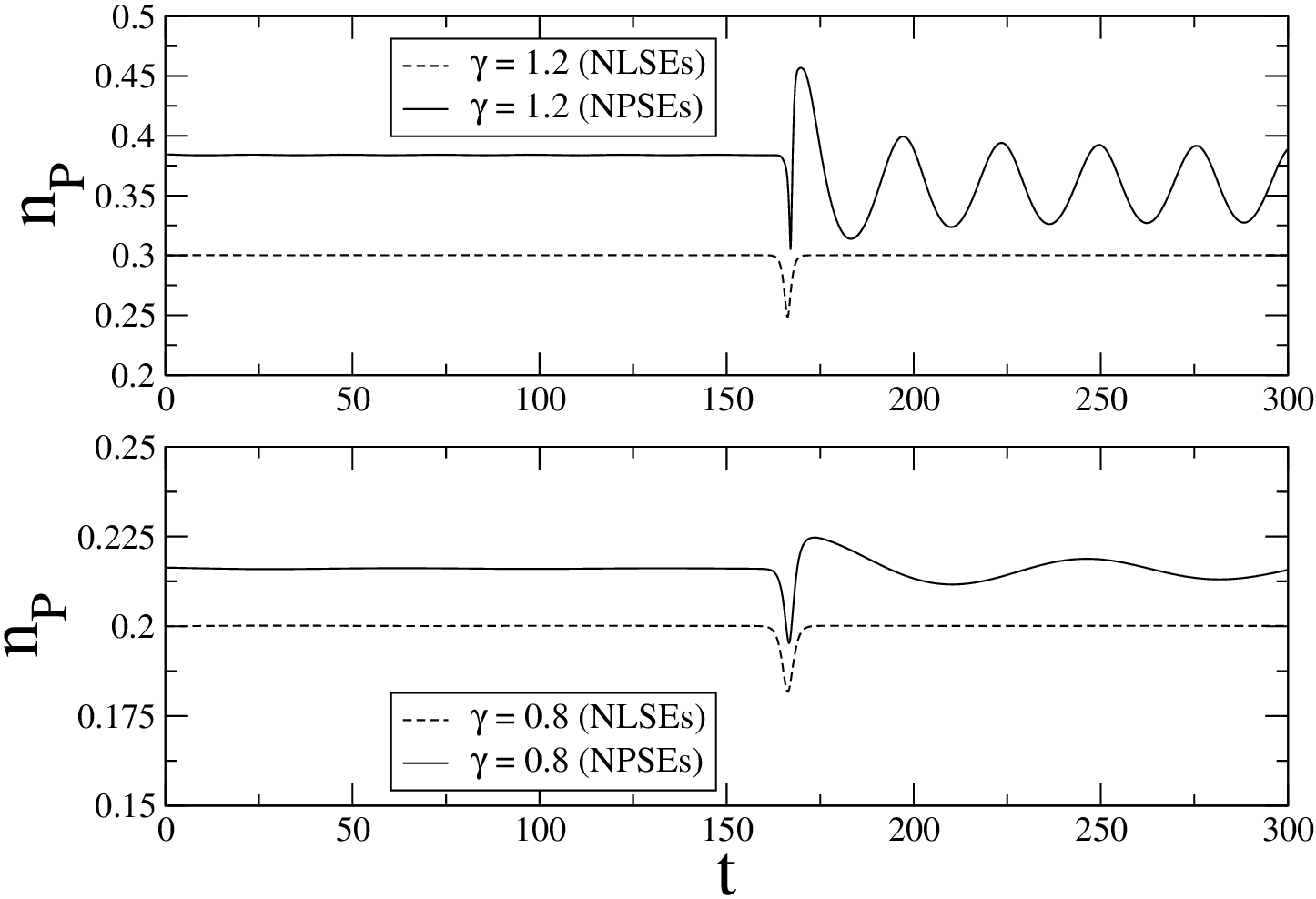}}
\caption{Peak density $n_{P}$ of the two colliding solitons as a function of
time ($t$), in the nonintegrable system of NPSEs, Eqs. (\protect\ref{final}%
), and in the integrable Manakov's system (alias NLSEs), Eqs. (\protect\ref%
{nlse}). In both cases, the initial velocity is $v=0.6$. }
\label{Fig1}
\end{figure}

As shown in Fig. 2, the amplitude and frequency ($\omega $) of the
oscillations excited by the collisions of solitons in the NPSE system grow
with interaction strength $\gamma $. On the contrary to that, the
simulations demonstrate that the amplitude and frequency of the intrinsic
oscillations \emph{do not }depend on initial velocity $v$ (see Fig. 5
below). The independence of $\omega \ $on $v$ is quite natural, as the
frequency is determined solely by the internal structure of the soliton.

\begin{figure}[tbp]
\centerline
{\includegraphics[height=2.35in,clip] {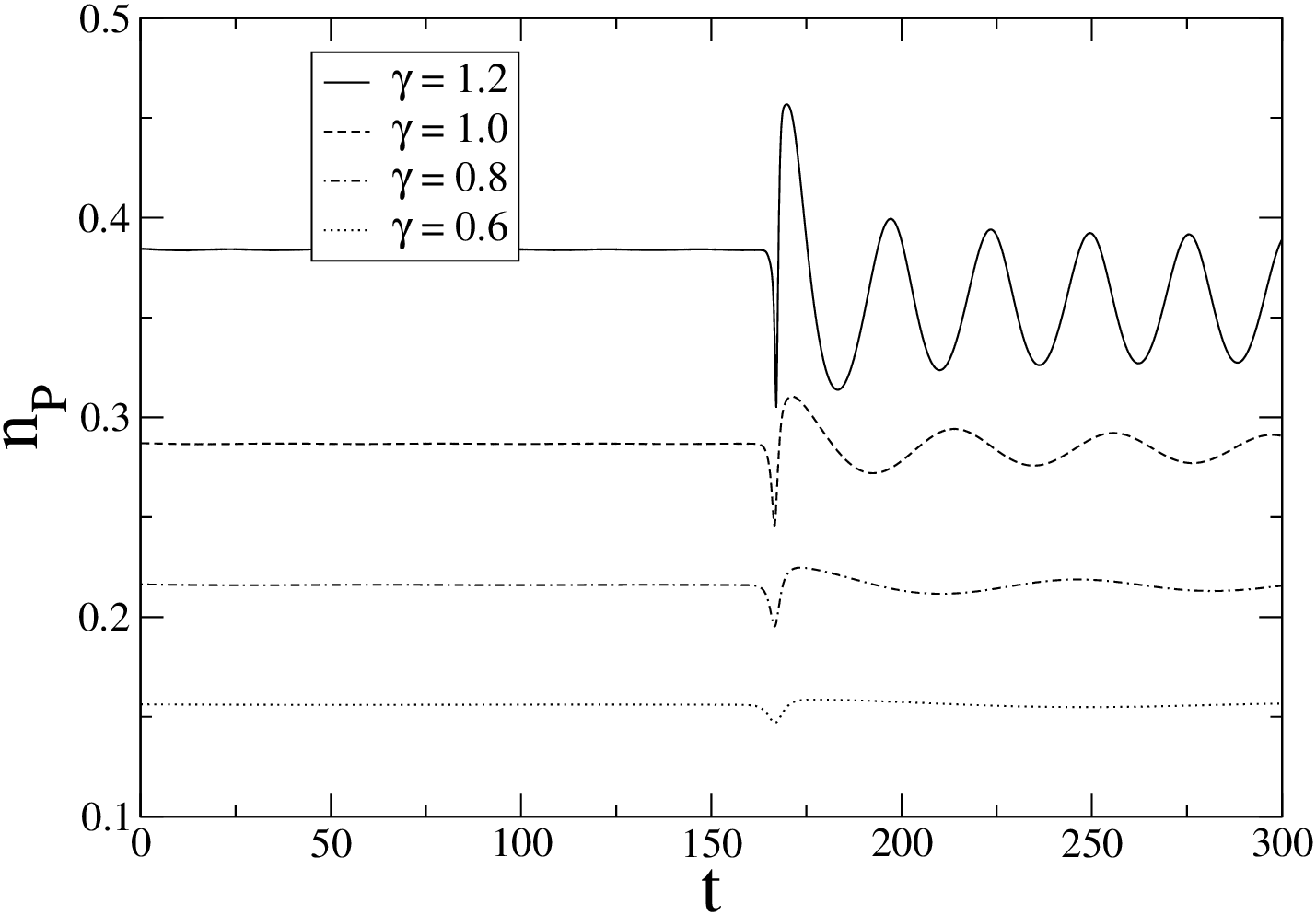}}
\caption{ Peak density $n_{P}$ of the two colliding solitons as a function
of time, found from numerical integration of the NPSEs, Eqs. (\protect\ref%
{final}), for different values of interaction strength $\protect\gamma $.
The initial velocity is $v=0.6$.}
\label{Fig2}
\end{figure}

In Fig. 3 we compare the intrinsic frequency, $\omega $, as found from the
direct simulations of the NPSEs, Eqs. (\ref{final}), to the frequency
calculated by means of the variational approach of Ref. \cite{sala2}. The
figure shows that the variational approximation somewhat overestimates both
the critical strength of longitudinal collapse, $\gamma_{c}$, and frequency $%
\omega $. For both quantities, the relative error is about $15\%$.

\begin{figure}[tbp]
\centerline
{\includegraphics[height=2.35in,clip] {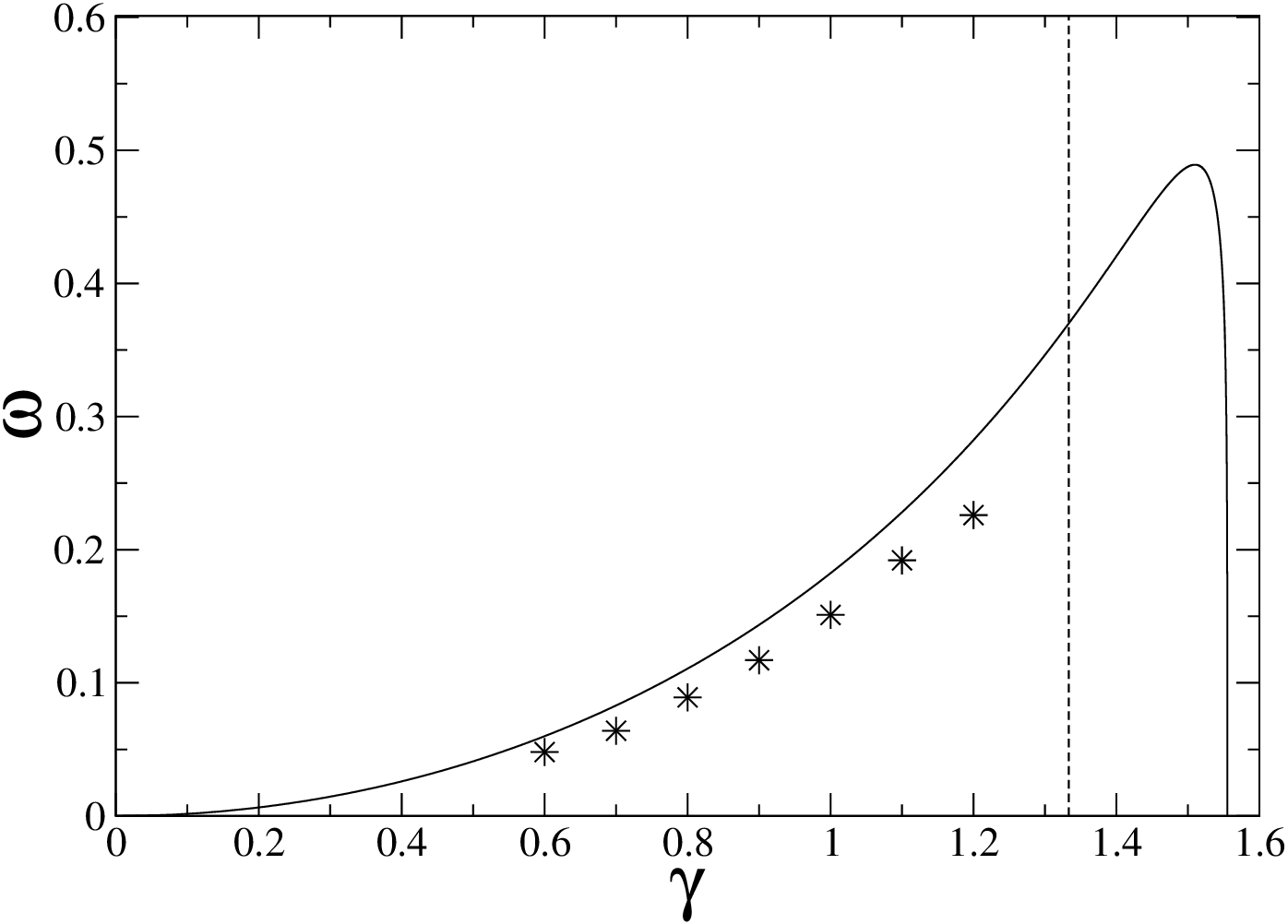}}
\caption{Frequency $\protect\omega $ of the intrinsic oscillations of the
two solitons as a function of interaction strength $\protect\gamma $.
Stars:\ $\protect\omega $ found from the numerical solution of NPSEs, Eqs. (%
\protect\ref{final}), as the frequency of intrinsic oscillations excited by
the collision between two solitons. Solid line: a result of the variational
approximation from Ref. \protect\cite{sala2}. The collapse point, found from
simulations of Eqs. (\protect\ref{final}), is indicated by the dashed
vertical line. The initial velocity is $v=0.8$.}
\label{Fig3}
\end{figure}

Collisions between solitons in NPSEs give rise to another noteworthy effect:
after the collision, a small part of the field, $\phi _{1}$, which
originally belonged to the first soliton remains trapped in the second
soliton, and vice versa, see Fig. 4. The effect is visible only for $\gamma
>1$, in the velocity interval of $0.4<v<0.7$ (outside this parameter range,
the effects takes place too but is very weak). Note that a similar
effect (``shadow formation") was observed in the model describing
the interaction of two polarizations of light in a nonlinear
optical fiber, which was based on a nonintegrable system of two
NLSEs with the cubic nonlinearity, see Refs. \cite{boris1,boris2}
and references therein. An explanation of the trapping effect is
based on the fact that the soliton in each field $\phi _{k}(z,t)$
($k=1,2$) supports not only the above-mentioned mode of intrinsic
vibrations in the same field, but also an external eigenmode of
small perturbations in the mate field, $\phi _{3-k}(z,t)$. The
latter mode is excited as a result of the collision \cite{boris1}.

\begin{figure}[tbp]
\centerline
{\includegraphics[height=2.35in,clip] {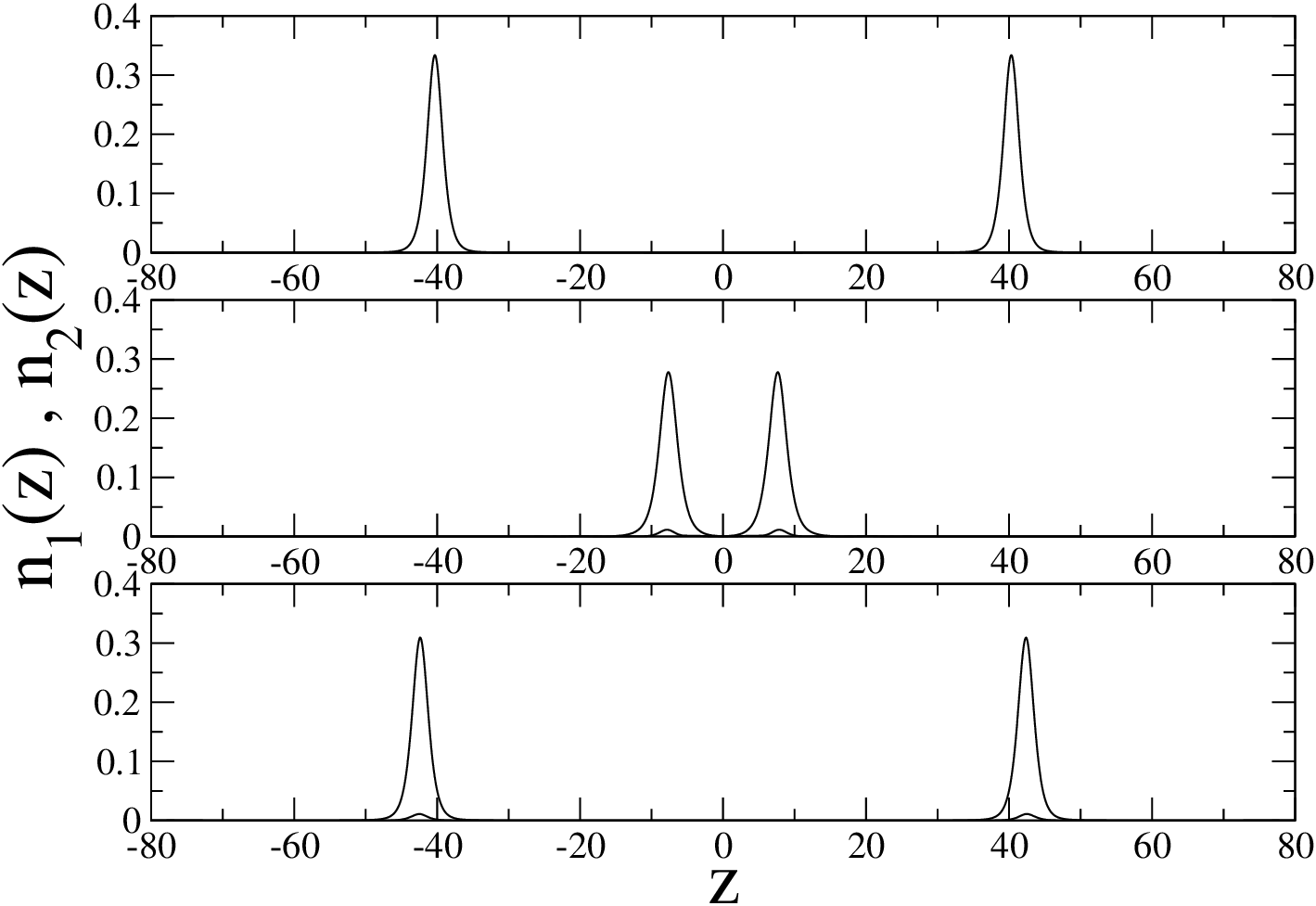}}
\caption{The trapping effect in the collision of two solitons: small parts
of fields $\protect\phi _{2}(z,t)$ and $\protect\phi _{1}(z,t)$ remain
bound, respectively, in the first and second soliton after the collision.
Top, central, and bottom panels display the density configurations, $%
n_{1,2}(z)\equiv \left\vert \protect\phi _{1,2}(z)\right\vert ^{2}$, at $%
t=100$, $180$, and $240$. The interaction strength is $\protect\gamma =1.2$,
and the initial velocity is $v=0.6$.}
\label{Fig4}
\end{figure}

The collisions feature a trend to be more inelastic at smaller velocities,
as illustrated by Fig. 5, which shows the maximum and minimum values of the
peak density, $n_{P}^{(M)}$ and $n_{P}^{(m)}$, in the oscillating solitons
emerging from the collision. The figure shows that, while $n_{P}^{(M)}$ does
not depend on initial velocity $v$, $n_{P}^{(m)}$ is smaller at smaller
velocities, which implies stronger inelasticity. The results are shown in
Fig. 5 for $\gamma =1$, and similar trends are found for $\gamma \leq 1$.
For completeness, in Fig. 5 we also plot frequency $\omega $ of the
intrinsic-mode excited by the collision, which confirms that $\omega $ does
not depend on $v$.

\begin{figure}[tbp]
\centerline
{\includegraphics[height=2.35in,clip] {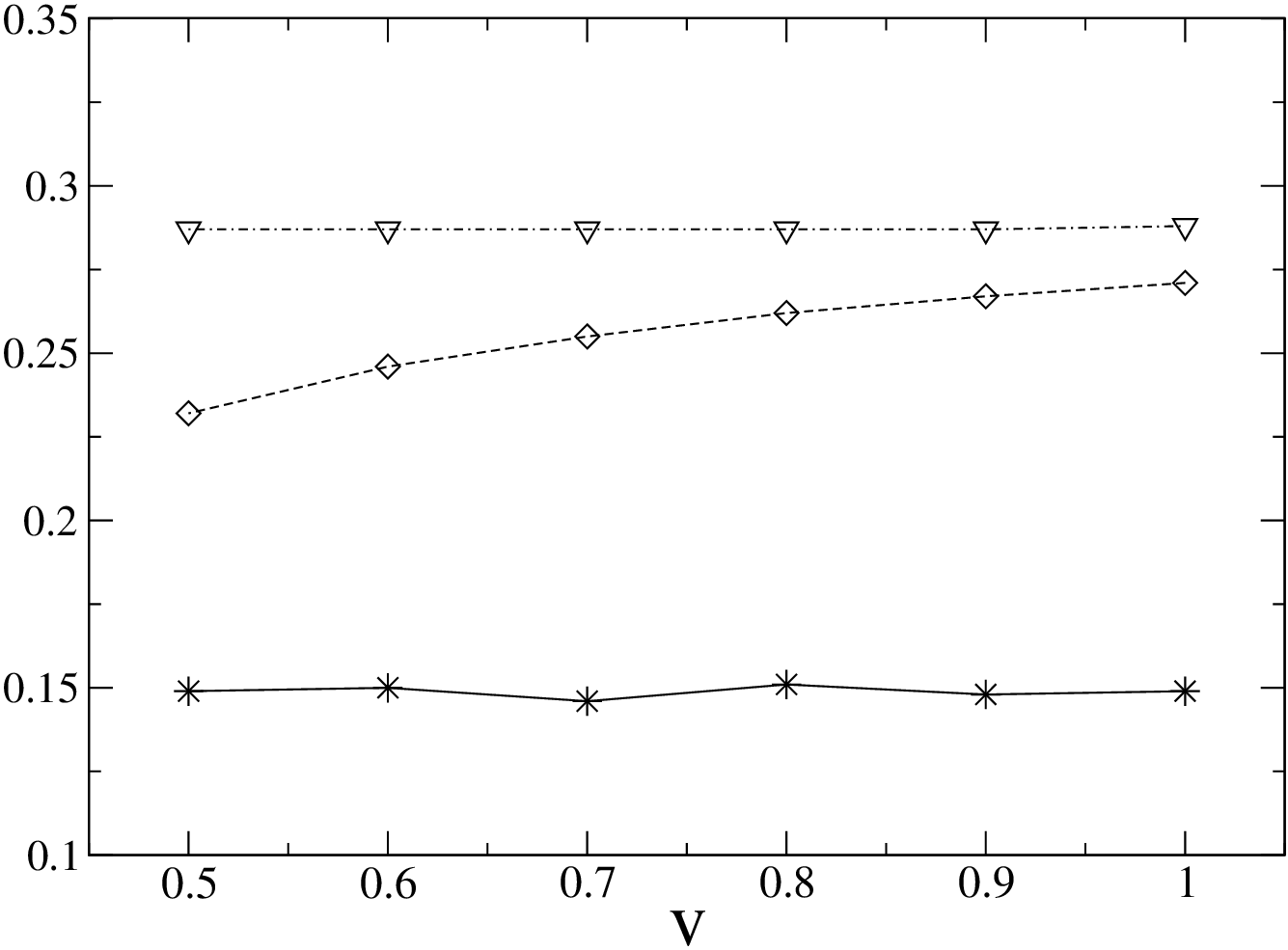}}
\caption{Dependence on the collision velocity $v$ of the maximum (triangles)
and minimum (rhombuses) values of the peak density, $n_{P}^{(M)}$ and $%
n_{P}^{(m)}$, in solitons disturbed by the collision. The respective
frequency of the intrinsic mode, $\protect\omega $, is shown by stars. The
interaction strength is $\protect\gamma =1$.}
\label{Fig. 5}
\end{figure}

An interesting issue is whether the collision may result in collapse. As
follows from Eqs. (\ref{final}), the collapse happens when condition $%
|g|(|f_{1}|^{2}+|f_{2}|^{2})=1$ takes place at some point. If the first
maximum of the peak density is achieved when the centers of the colliding
solitons nearly coincide, this condition can be estimated as $%
n_{P}^{(M)}\simeq 1/(2\gamma )$. While Fig. 5 shows that $n_{P}^{(M)}$ does
not depend on collision velocity $v$ at $\gamma \leq 1$, it depends on $v$
at $\gamma >1$, and the collapse can thus been reached by increasing $v$. In
particular, Fig. 6 shows that, for $\gamma =1.2$, the maximum value of peak
density, $n_{P}(t)$, grows with $v$, and the collapse takes place at $v=0.9$%
. Note that collapse induced by the collision between two solitons in a
single NPSE was reported in Ref. \cite{sala2}, but in that case the onset of
the collapse did not depend on the initial velocity.

\begin{figure}[tbp]
\centerline
{\includegraphics[height=2.35in,clip] {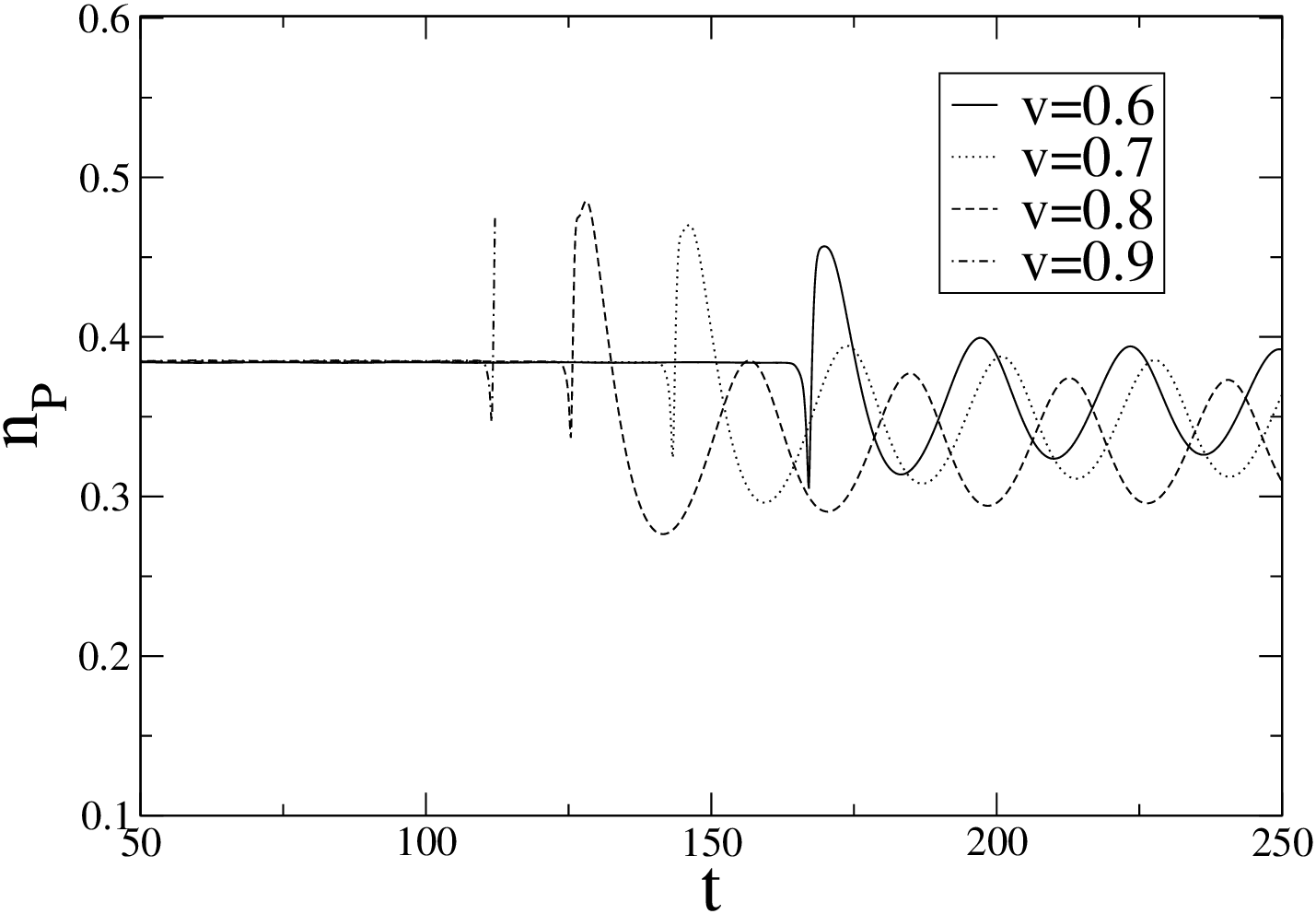}}
\caption{Peak density $n_{P}$ of the two colliding solitons as a function of
time for different values of initial velocity $v$. The collapse is induced
by the collision at $v=0.9$ (the corresponding curve shoots up vertically at
the collapse moment, $t=t_{c}\approx 110$). The interaction strength is $%
\protect\gamma =1.2$, and the initial separation is $z_{0}=200$.}
\label{Fig6}
\end{figure}

\section{Conclusions}

In this work, we have derived a system of one-dimensional coupled
nonpolynomial Schr\"{o}dinger equations (NPSEs) for longitudinal wave
functions of two components in a binary BEC, in the case of attraction
between the atoms. The system was derived by means of the variational
approximation, starting from the coupled 3D Gross-Pitaevskii equations for
the two species and assuming (as usual) the factorization of 3D wave
functions into products of the strongly confined transverse and slowly
varying longitudinal ones. The system was cast in a tractable form under a
natural symmetry constraint. Then, a family of two-component (vector)
soliton solutions was obtained, and collisions between orthogonal solitons
(each belonging to one component only) were studied in detail by dint of
systematic numerical simulations. It was found that the collisions are
inelastic. They lead to strong excitation of intrinsic oscillations in the
solitons emerging from the collision, and to formation of a small orthogonal
component (``shadow") in each soliton. Eventually, the collision may
initiate collapse of the solitons, depending on their mass and velocities.

\section*{Acknowledgements}

L.S. thanks Alberto Parola and Luciano Reatto for many discussions and
suggestions. The work of B.A.M. was supported, in a part, by the Israel
Science Foundation through the Center-of-Excellence grant No. 8006/03.

\end{document}